\documentclass{article}

\def\xxinput#1{\input#1}

\xxinput{vsolj02.sty}

\usepackage[dvipdfmx]{graphicx}
\usepackage[comma,colon]{natbib}

\usepackage[OT2,T1]{fontenc}

\def\cite{\citealt}
\setcitestyle{aysep={}}

\newcounter{author}
\def\altaffilmark#1{$^{#1}$}
\def\altaffiltext#1{$^{#1}$\,}

\def\authorcount#1#2{{\refstepcounter{author}\label{#1}
                     \altaffiltext{\ref{#1}}{#2}}}

\begin{document}

\begin{center}

\title{MGAB-V240: 23-min AM CVn star showing both 12-d supercycle and standstills}

\author{
        Taichi~Kato\altaffilmark{\ref{affil:Kyoto}}}
\email{tkato@kusastro.kyoto-u.ac.jp}

\authorcount{affil:Kyoto}{
     Department of Astronomy, Kyoto University, Sakyo-ku,
     Kyoto 606-8502, Japan}

\end{center}

\begin{abstract}
\xxinput{abst.inc}
\end{abstract}

\begin{figure*}
\begin{center}
\includegraphics[width=14cm]{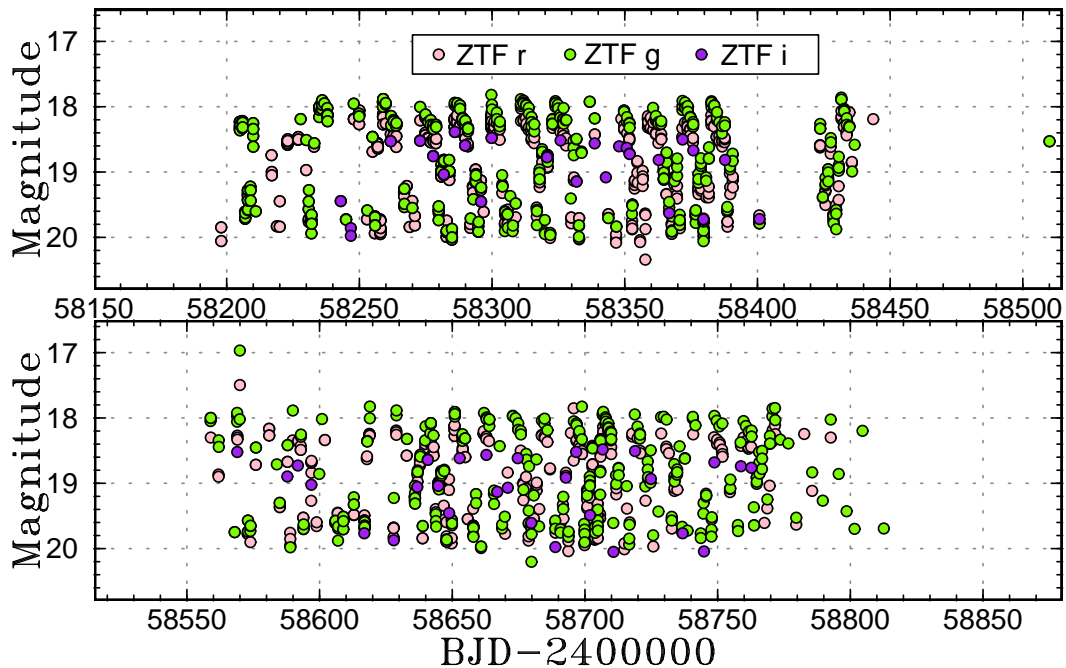}
\caption{
   Light curve of MGAB-V240 in 2018--2019.
   The object showed a regular pattern of outbursts recurring
   with a period of $\sim$12~d.  They turned out to be
   superoutbursts, not SS Cyg-type outbursts (see text).
}
\label{fig:lc1}
\end{center}
\end{figure*}

\begin{figure*}
\begin{center}
\includegraphics[width=14cm]{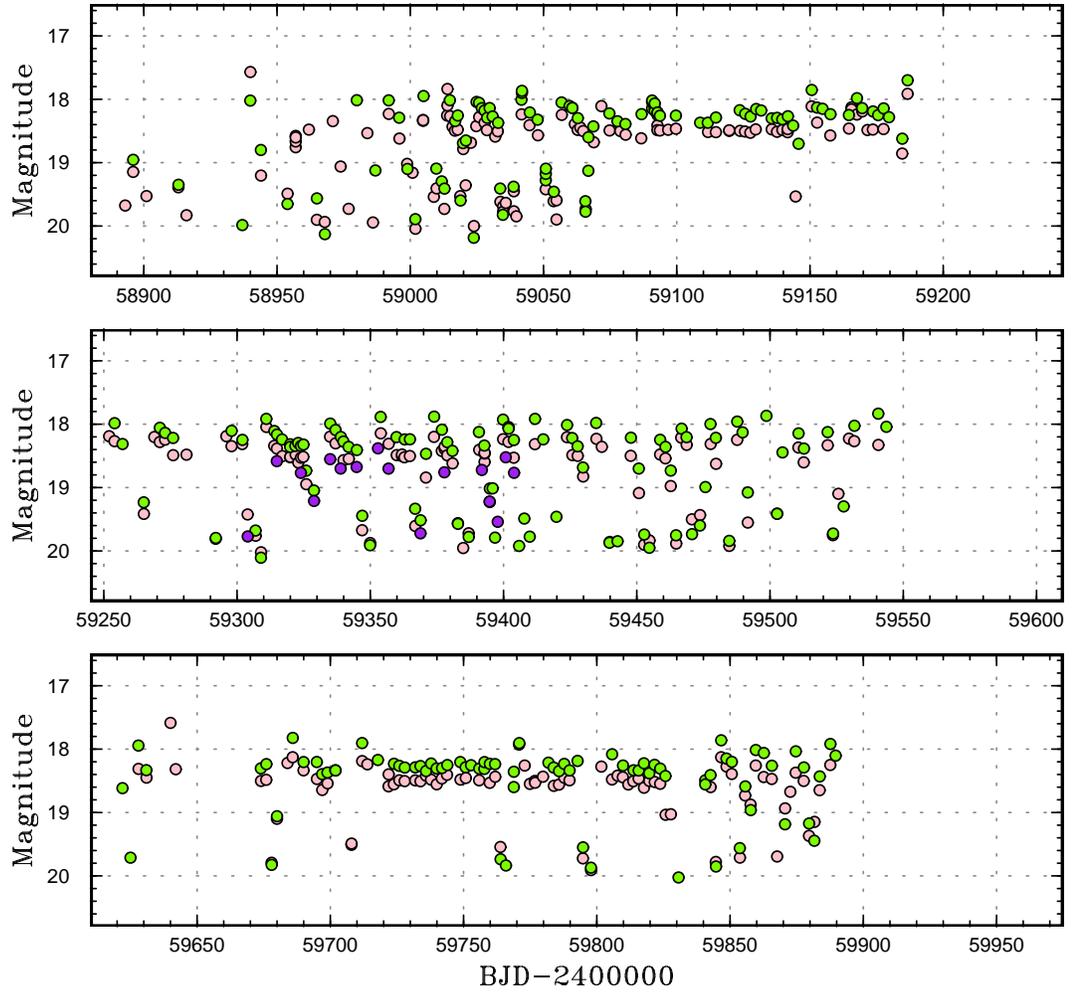}
\caption{
   Light curve of MGAB-V240 in 2020--2022.
   The symbols are the same as in figure \ref{fig:lc1}.
   Standstills were present in addition to outbursts.
}
\label{fig:lc2}
\end{center}
\end{figure*}

\begin{figure*}
\begin{center}
\includegraphics[width=14cm]{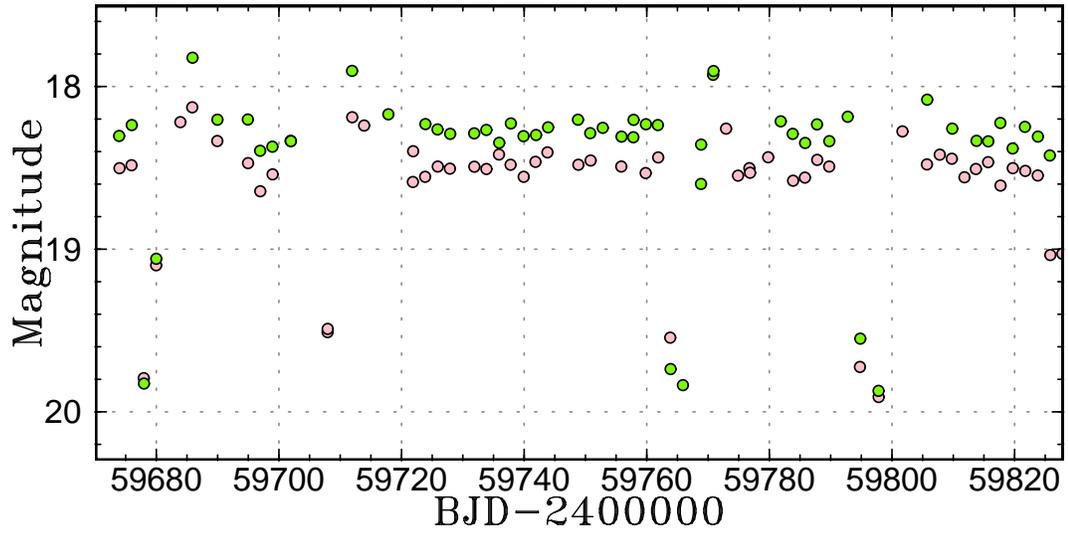}
\caption{
   Enlargement of the 2022 standstill of MGAB-V240.
   The symbols are the same as in figure \ref{fig:lc1}.
   Sudden drops from the standstill were recorded.
   The object usually brightened after these drops.
}
\label{fig:lc4}
\end{center}
\end{figure*}

   MGAB-V240 was discovered as a faint SS Cyg-type dwarf nova
with frequent outbursts.\footnote{
   $<$https://www.aavso.org/vsx/index.php?view=detail.top\&oid=702830$>$.
}  The object had also been selected as a candidate RR Lyr star
(PS1-3PI J185529.82$+$323017.8, \cite{ses17PS1RR}).\footnote{
   Although the name PS1-3PI J185529.82$+$323017.8 would be adequate
   considering the priority in discovery, I use MGAB-V240 in this paper
   because of its brevity.
}
I used Zwicky Transient Facility (ZTF: \cite{ZTF})\footnote{
   The ZTF data can be obtained from IRSA
$<$https://irsa.ipac.caltech.edu/Missions/ztf.html$>$
using the interface
$<$https://irsa.ipac.caltech.edu/docs/program\_interface/ztf\_api.html$>$
or using a wrapper of the above IRSA API
$<$https://github.com/MickaelRigault/ztfquery$>$.
} data and found that this object showed standstills in
2020 and 2022.  The long-term light curves containing
a regularly outbursting part (2018--2019) and a state
with standstills and outbursts (2020--2022) are shown in figures
\ref{fig:lc1} and \ref{fig:lc2}, respectively.
A na\"{\i}ve look at these figures would simply re-classify
the object as an Z Cam star
[for cataclysmic variables and their subclasses, see
e.g., \citet{war95book}].
I, however, noticed that the object showed sudden drops
during the 2022 standstill (figure \ref{fig:lc4}),
which is unusual for a Z Cam-type dwarf nova
(vsnet-chat 9317).\footnote{
   $<$http://ooruri.kusastro.kyoto-u.ac.jp/mailarchive/vsnet-chat/9317$>$.
}
One or two drops were also recorded during the 2020 standstill.
The fading rates of these sudden drops were sometimes
close to 2~mag d$^{-1}$, whose large value is one of the signatures
of AM CVn-type outbursts \citep{kat21newAMCVn}
[for a review of AM CVn stars, see e.g., \citet{sol10amcvnreview}].

\begin{figure*}
\begin{center}
\includegraphics[width=14cm]{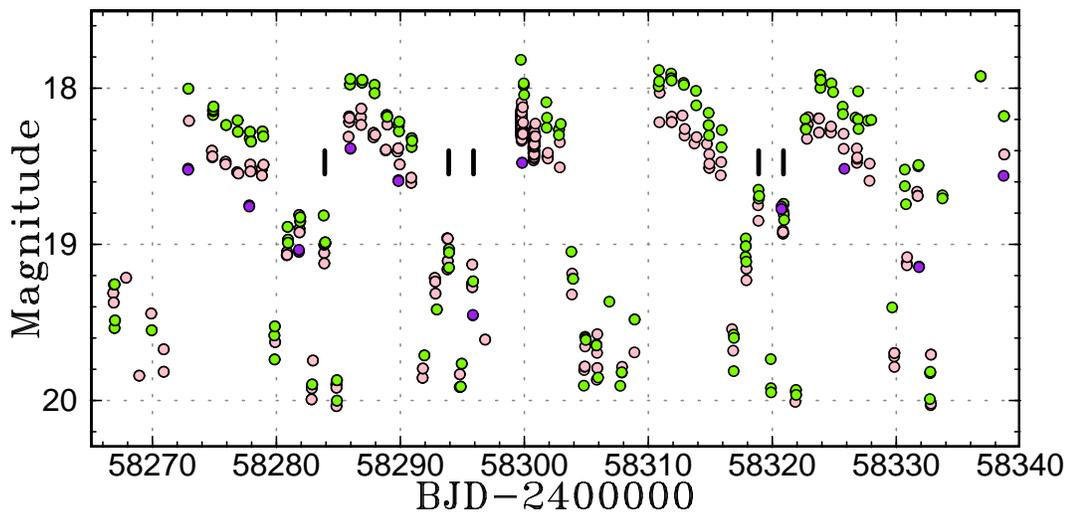}
\caption{
   Enlargement of the 2018 outbursts of MGAB-V240.
   The symbols are the same as in figure \ref{fig:lc1}.
   In addition to relatively regular long outbursts, short
   outbursts (with tick marks) were also present.
   Some outbursts between the long ones apparently had durations
   more than 1~d and they were not marked by the ticks.
}
\label{fig:lcso}
\end{center}
\end{figure*}

\begin{figure*}
\begin{center}
\includegraphics[width=14cm]{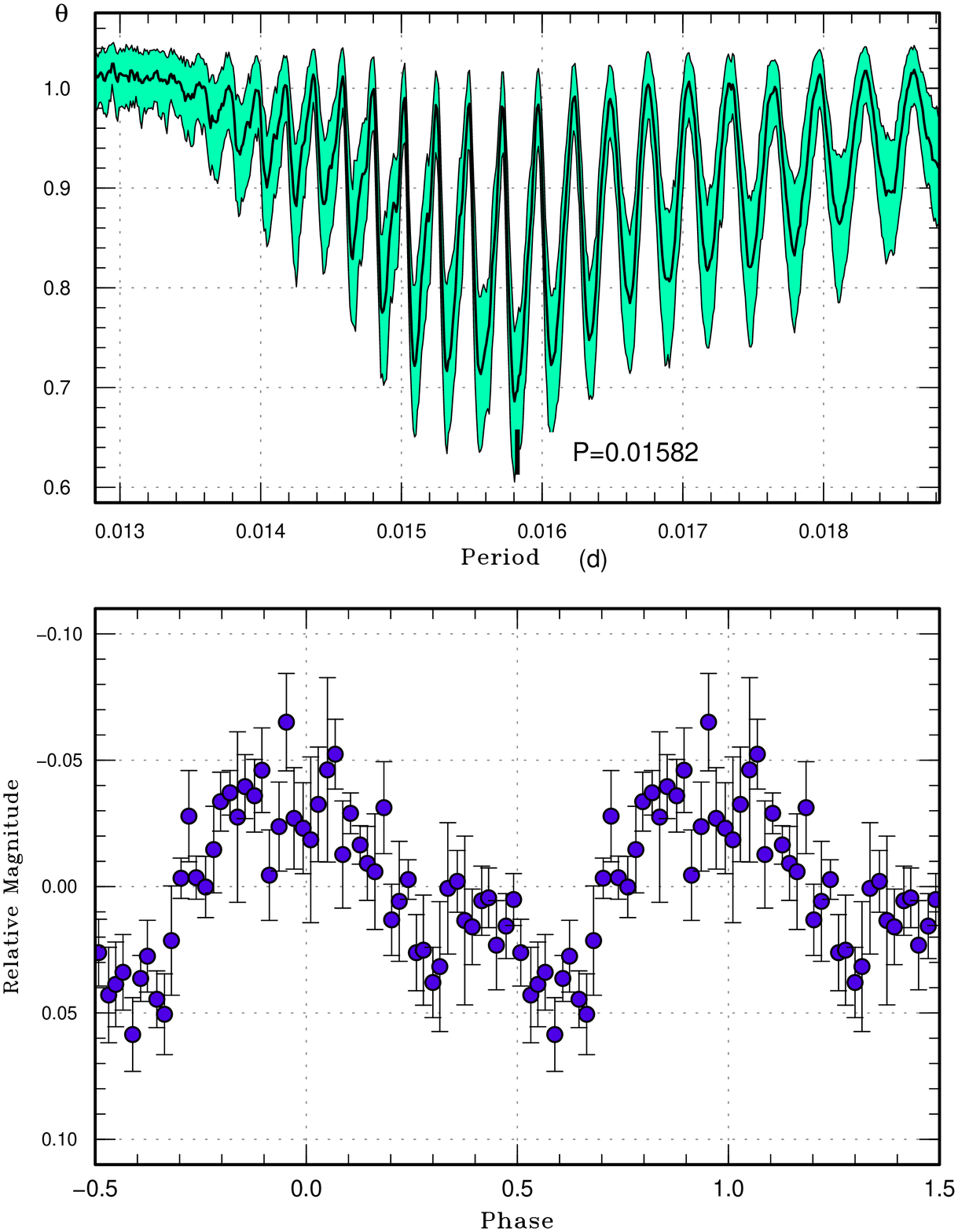}
\caption{
   Superhumps on 2018 June 30--July 1 recorded by
   ZTF time-resolved photometry.
   (Upper): PDM analysis.  The bootstrap result using
   randomly contain 50\% of observations is shown as
   a form of 90\% confidence intervals in the resultant 
   $\theta$ statistics.
   (Lower): Phase plot.
}
\label{fig:sh}
\end{center}
\end{figure*}

   Upon a closer look at the ZTF light curve of
the regularly outbursting part, I found short outbursts
between long outbursts (figure \ref{fig:lcso}).
This is a clear indication of an SU UMa-type supercycle
(long outbursts and short outbursts between them).
Superhumps with a period of 0.015824(9)~d were indeed detected
from the ZTF time-resolved data during one of long outbursts
(figure \ref{fig:sh}).
The error of the period was estimated by
the methods of \citet{fer89error} and \citet{Pdot2}.
The overall light curve of the 2018--2019 season
resembles that of ASASSN-14cc \citep{kat15asassn14cc}.
\citet{kat15asassn14cc} detected a supercycle of 21--33~d
together with superhumps with period of 0.01560--0.01562~d
by a network of ground-based small telescopes aiming at
ASASSN-14cc under a VSNET \citep{VSNET} campaign.
This period was confirmed by TESS photometry \citep{pic21amcvns}.
\citet{kat15asassn14cc} suggested that ASASSN-14cc
showed a supercycle similar to the hydrogen-rich system
RZ LMi.  RZ LMi typically has a supercycle of 19~d
\citep{rob95eruma,nog95rzlmi,ole08rzlmi}, but showed
a short standstill in 2016 \citep{kat16rzlmi}.
Such a short supercycle could not be naturally reproduced
\citep{osa95eruma,osa95rzlmi} by the thermal-tidal instability
model \citep{osa89suuma,osa96review}, which successfully
explained the supercycles of most SU UMa stars.
\citet{osa95rzlmi} explained the short supercycle
by artificially quenching the superoutburst when
the accretion disk is still large.  This treatment led to
an idea of decoupling between the thermal and tidal
instabilities \citep{hel01eruma}.

   The present observations of MGAB-V240 make this object
as a perfect helium analog (although no spectrum has been
obtained, the short superhump period is an unambiguous signature
of an AM CVn-type object) of the hydrogen-rich RZ LMi:
a short-period supercycle and standstills.
True standstills in AM CVn stars have been very rare
and this object becomes the second well-established example
after CR Boo \citep{kat23crboo}.  The suggested type is
SU UMa(ER UMa)+Z Cam+AM CVn.

   In AM CVn stars, an RZ LMi-like short supercycle would be
more easily achieved than in hydrogen-rich systems
for two reasons: (1) Helium disks require a higher temperature
to maintain the hot state and a cooling wave starts more
easily than in hydrogen-rich systems.
(2) AM CVn stars have (usually) lower mass ratios than in
hydrogen-rich systems, which would make decoupling between
the thermal and tidal instabilities easier to happen
[For the disk-instability model of AM CVn stars,
see \citet{tsu97amcvn,sol10amcvnreview,kot12amcvnoutburst}].
Drops from standstills may be caused by the difficulty
(relative to hydrogen-rich systems) in maintaining
the hot state.  In contrast to most (hydrogen-rich) Z Cam
stars, these drops were often followed by short brightening
(after the initial three drops in figure \ref{fig:lcso}).
This phenomenon can be interpreted as follows:
The mass transfer from the secondary and the mass accretion
to the primary are balanced during the standstill.
Once a cooling wave starts, the mass accretion to the primary
decreases and the mass accumulates (if the mass-transfer rate
is constant) during these drops in the outer part of the disk, and
this extra mass causes brightening when the disk becomes hot again.
The phenomenon observed in standstills of MGAB-V240 excludes
the possibility of a temporary decrease of the mass-transfer rate
as the cause of the drops; rather the constant mass-transfer rate
is favored to explain brightening after the drops.

   The superhump period of 0.015824~d, which is usually $\sim$1\%
longer than the orbital period ($P_{\rm orb}$) in AM CVn stars,
in MGAB-V240 is between the thermally unstable dwarf nova-type CR Boo
($P_{\rm orb}$=0.017029~d: \cite{pro97crboo}) and
the thermally stable novalike-type HP Lib
($P_{\rm orb}$=0.012763~d: \cite{pat02hplib,roe07hplibv803cen})
among AM CVn stars and follows the activity sequence expected by
the disk instability theory and the evolutionary sequence
of AM CVn stars \citep{tsu97amcvn,kot12amcvnoutburst}.

\section*{Acknowledgements}

This work was supported by JSPS KAKENHI Grant Number 21K03616.

I am grateful to Naoto Kojiguchi for helping downloading the ZTF data
and the ZTF team for making their data
available to the public.

Based on observations obtained with the Samuel Oschin 48-inch
Telescope at the Palomar Observatory as part of
the Zwicky Transient Facility project. ZTF is supported by
the National Science Foundation under Grant No. AST-1440341
and a collaboration including Caltech, IPAC, 
the Weizmann Institute for Science, the Oskar Klein Center
at Stockholm University, the University of Maryland,
the University of Washington, Deutsches Elektronen-Synchrotron
and Humboldt University, Los Alamos National Laboratories, 
the TANGO Consortium of Taiwan, the University of 
Wisconsin at Milwaukee, and Lawrence Berkeley National Laboratories.
Operations are conducted by COO, IPAC, and UW.

The ztfquery code was funded by the European Research Council
(ERC) under the European Union's Horizon 2020 research and 
innovation programme (grant agreement n$^{\circ}$759194
-- USNAC, PI: Rigault).

\section*{List of objects in this paper}
\xxinput{objlist.inc}

\section*{References}

We provide two forms of the references section (for ADS
and as published) so that the references can be easily
incorporated into ADS.

\newcommand{\noop}[1]{}\newcommand{\hyphalt}{-}

\renewcommand\refname{\textbf{References (for ADS)}}

\xxinput{mgabv240aph.bbl}

\renewcommand\refname{\textbf{References (as published)}}

\xxinput{mgabv240.bbl.vsolj}

\end{document}